 \documentclass[pmlr,twocolumn,10pt]{jmlr} 

\usepackage{booktabs}
\usepackage{siunitx}

\usepackage[utf8]{inputenc} 
\usepackage[T1]{fontenc}    
\usepackage{hyperref}       
\usepackage{url}            
\usepackage{booktabs}       
\usepackage{amsfonts}       
\usepackage{nicefrac}       
\usepackage{microtype}      
\usepackage{xcolor}         
\usepackage{graphicx}       
\usepackage[none]{hyphenat}

\usepackage{mathtools}
\usepackage[para]{footmisc}
\usepackage{refcount}
\usepackage{threeparttable}     
\usepackage{amssymb}            
\usepackage{float}          
\usepackage{subfiles}

\usepackage{ragged2e}

\usepackage{multirow}

\newcommand{\mycomment}[1]{}

\theorembodyfont{\upshape}
\theoremheaderfont{\scshape}
\theorempostheader{:}
\theoremsep{\newline}

\jmlrvolume{LEAVE UNSET}
\jmlryear{2023}
\jmlrsubmitted{LEAVE UNSET}
\jmlrpublished{LEAVE UNSET}
\jmlrworkshop{Machine Learning for Health (ML4H) 2023} 

 \title[Enhancing Transformer-Based Image Segmentation for Breast Cancer Diagnosis]{Enhancing Transformer-Based Segmentation for Breast Cancer Diagnosis using Auto-Augmentation and Search Optimisation Techniques}

\author{
\Name{Leon Hamnett}
\Email{L.s.hamnett@gmail.com} \\
\addr{CARESAI\textsuperscript{1}}
\AND
\Name{Mary Adewunmi}
\Email{mary.adewunmi@utas.edu.au}\\
\addr{University of Tasmania\textsuperscript{2}, CARESAI\textsuperscript{1}}
\AND
\Name{Modinat Abayomi}
\Email{modinat.abayomi@bc.edu} \\
\addr{Boston College\textsuperscript{3}, CARESAI\textsuperscript{1}}
\AND
\Name{Kayode Raheem}
\Email{kayoderaheem.y@gmail.com} \\
\addr{COMSATS University
Islamabad\textsuperscript{4}, CARESAI\textsuperscript{1}}
\AND
\Name{Fahad Ahmed}
\Email{fahad.ahmed@tib.eu} \\ 
\addr{Leibniz University\textsuperscript{5},  CARESAI\textsuperscript{1}}
}

\begin{document}

\maketitle

\begin{abstract}
Breast cancer remains a critical global health challenge, necessitating early and accurate detection for effective treatment. This paper introduces a methodology that combines automated image augmentation selection (RandAugment) with search optimisation strategies (Tree-based Parzen Estimator) to identify optimal values for the number of image augmentations and the magnitude of their associated augmentation parameters, leading to enhanced segmentation performance. We empirically validate our approach on breast cancer histology slides, focusing on the segmentation of cancer cells. A comparative analysis of state-of-the-art transformer-based segmentation models is conducted, including SegFormer, PoolFormer, and MaskFormer models, to establish a comprehensive baseline, before applying the augmentation methodology. Our results show that the proposed methodology leads to segmentation models that are more resilient to variations in histology slides whilst maintaining high levels of segmentation performance, and show improved segmentation of the tumour class when compared to previous research. Our best result after applying the augmentations is a Dice Score of 84.08 and an IoU score of 72.54 when segmenting the tumour class. The primary contribution of this paper is the development of a methodology that enhances segmentation performance while ensuring model robustness to data variances. This has significant implications for medical practitioners, enabling the development of more effective machine learning models for clinical applications to identify breast cancer cells from histology slides. Furthermore, the codebase accompanying this research will be released upon publication. This will facilitate further research and application development based on our methodology, thereby amplifying its impact.\\

\end{abstract}
\maketitle

\footnotetext[1]{Cancer Research with AI - CARESAI}
\footnotetext[2]{College of Health \& Medicine, School of Medicine, University of Tasmania (UTAS), Hobart, Australia}
\footnotetext[3]{Boston College, Department of Biology, USA}
\footnotetext[4]{Chemical Biology and Drug Discovery Lab, Department of Bioscience, COMSATS University Islamabad, Pakistan}
\footnotetext[5]{TiB - Leibniz Information Center for Science and Technology, Hannover, Germany}

\section{Introduction}
Breast cancer is a significant global health issue, accounting for a staggering one in eight of all cancer diagnoses worldwide, and the impact of breast cancer will increase significantly by the year 2040 \citep{cancer_net}. Breast cancer involves the growth of malignant cells forming a lesion within the breast tissue. Early detection and treatment of these malignant cells has been shown to have a significant impact on successful treatment of the cancer and patient outcomes \citep{ott2009importance}. Histopathological analysis (examining cells and tissues under a microscope) \citep{roy2020macroscopic} is considered the gold standard for detecting most cancers, as this reveals the microscopic tissue structure, and can confirm the benign or malignant nature of a suspected tumour \citep{kumar2017robbins, mills2012sternberg}. 

However, the manual evaluation of histopathological slides is time-consuming and challenging due to the complex appearance of tissues and cells, inconsistent staining and variation in illumination between different laboratories, overlapped and clustered nuclei, and poorly fixed tissues \citep{aksac2020cactus}. An accurate diagnosis of tumorous cells depends on the pathologist's experience and skill \citep{das2020computer}, however, many developing countries suffer from a shortage of trained pathologists \citep{ziegenhorn2020breast,sayed2015providing} and this lack of skilled pathologists presents an additional challenge for early detection and successful treatment of breast cancer in these countries \citep{yan2020breast}. There has been growing interest in the development of Computer-aided diagnostic systems (CAD) to increase the efficiency with which a pathologist can examine histopathology slides. CAD systems for histopathological slides typically involve image preprocessing, segmentation, feature extraction, and classification of slide images  \citep{he2012histology, demir2005automated}. CAD systems have begun to leverage machine learning and deep learning methods, particularly neural networks, to enhance the accuracy and efficiency of slide analysis \citep{madabhushi2016image, holzinger2020artificial,cohen2020artificial}. Image segmentation is an essential component of many computer vision and medical imaging systems, and this involves partitioning images into different classes and predicting pixel-level classification by assigning each pixel to a class category \citep{minaee2022image}, e.g. tumorous cells and healthy cells. Image segmentation models have recently shown remarkable performance improvements and are increasingly used across many medical fields \citep{minaee2022seg_review, moorthy2022medsurvey}. Applying these models to histopathology images aids in identifying potential Regions of Interest (ROIs) \citep{rashmi2021breast}, highlighting possible tumorous cells for review by a trained pathologist and speeding up pathologist efficiency. However, the time and effort required to create a segmentation dataset from breast cancer histology slides may have limited the adoption of image segmentation in the area of breast cancer research. 

\begin{figure}[h]
\vspace*{-0.1cm}
\label{figure:wsi_examples}
\centering
\includegraphics[width=\linewidth,height=2.0cm]{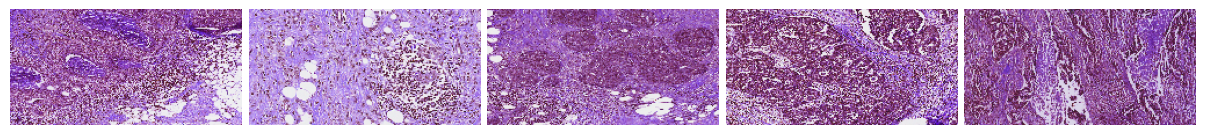}
\vspace*{-0.8cm}
\caption{Examples of patches taken from breast cancer histology slides}
\vspace*{-0.6cm}
\end{figure}

We see from existing research \citep{krithiga2020breast}, that image segmentation has successfully been applied to a number of other areas of cancer research, including lymphoma segmentation from histological images \citep{tosta2017segmentation}, nucleus detection in breast cancer histology slides \citep{lakshmanan2018nucleus}, renal cell cancer segmentation \citep{albayrak2018automatic}, and colon cancer cell segmentation \citep{jia2017constrained}. A recent study leveraged a crowd-sourced dataset to train a VGG16-FCN8 segmentation model for tumor segmentation in breast histology slides, achieving a significant Dice Score of 85.1 for the cancer class \citep{amgad2019structured}.  
Recently, across many computer vision applications, transformer-based models have shown increased research interest \citep{paperswithcode_vit} and seem to show better performance than traditional CNNs for a wide variety of computer vision tasks \citep{han2023survey_vision}. An extremely limited amount of previous research has been undertaken into the task of using transformer-based models to segment breast cancer histology images. Recent research \citep{lu2022automatic} has shown that the use of transformer-based architectures is able to achieve significant results for a similar segmentation task, by using a dual-path transformer model, achieving a mean intersect over union (mIoU) score of 68.12 when segmenting the cancer class. We note that, although impressive results have been achieved on similar tasks for segmenting cancer cells from histology slides \citep{lu2022automatic,amgad2019structured}, further increases in segmentation performance may still be possible by implementing a more optimal choice of image augmentations and associated augmentation parameters \citep{tellez2019_data_aug_hist}. There exists a wide range of possible image augmentations \citep{alomar2023data_aug}  which were not explored in the closest related works \citep{amgad2019structured,lu2022automatic}. For example, in one related paper \citep{amgad2019structured}, only a limited selection of augmentations such as colour normalisation, shift, and crop transforms were implemented. And in another related paper \citep{lu2022automatic}, only rotational augmentations were used. Applying a diverse set of image augmentations is key to making segmentation models robust to changes in input images and maintaining the ability to accurately identify potential areas of tumorous cells. This robustness is particularly important for models detecting cancerous cells in histology images as histology slides may vary considerably between different labs and even within the same lab due to different technicians' slide-staining processes and variation between slide-scanning machines \citep{bancroft2012hematoxylins}.  The authors hypothesise that a combination of state-of-the-art segmentation models, particularly transformer-based architectures, with a comprehensive and optimised selection of image augmentations, can further enhance the accuracy of cancer cell identification from breast cancer histology slides. To navigate the complex task of selecting optimal image augmentations and associated parameters, we propose a novel method to find optimal augmentation parameters which combines the RandAugment auto-augmentation strategy with a tree-structured parzen estimator optimisation strategy. 

\subsection{Research Questions}
The following research questions gave rise to the project implementation:\\

Research Question 1:\\
Which image segmentation transformer model gives the best performance for identifying tumorous regions from histopathology slides? This allows us to start from a high-performing baseline and iterate further for better performance.\\

Research Question 2:\\
Can a set of optimal image augmentations be discovered that improve segmentation metrics and make more robust models? By applying a number of different augmentations to training images, the model becomes more robust to dataset-specific features and is better able to generalise when making predictions on unseen data.\\

General research framework:\\
To address the above research questions, firstly the performance of transformer-based models for segmenting cancerous cells is assessed without image augmentation. Then the optimal number and magnitude of image augmentations are found which are then applied to the best-performing models. This research provides a framework for selecting optimum image augmentations, a valuable resource for machine learning and medical researchers.

\section{Methodology}

Our general approach is to specify a dataset, explore model selection and then apply the image augmentation optimisation process (including defining an objective function related to improving segmentation performance as well as defining a parameter search space) and then combining the best augmentations and models and performing cross-validation. 

\subsection{Dataset preparation and preprocessing}

\begin{figure}[h]
\vspace{0.05cm}
\setlength{\belowcaptionskip}{-2pt}
\centering
\label{figure:seg_mask_examples}
\includegraphics[width=1.0\linewidth,height=2.5cm,keepaspectratio]{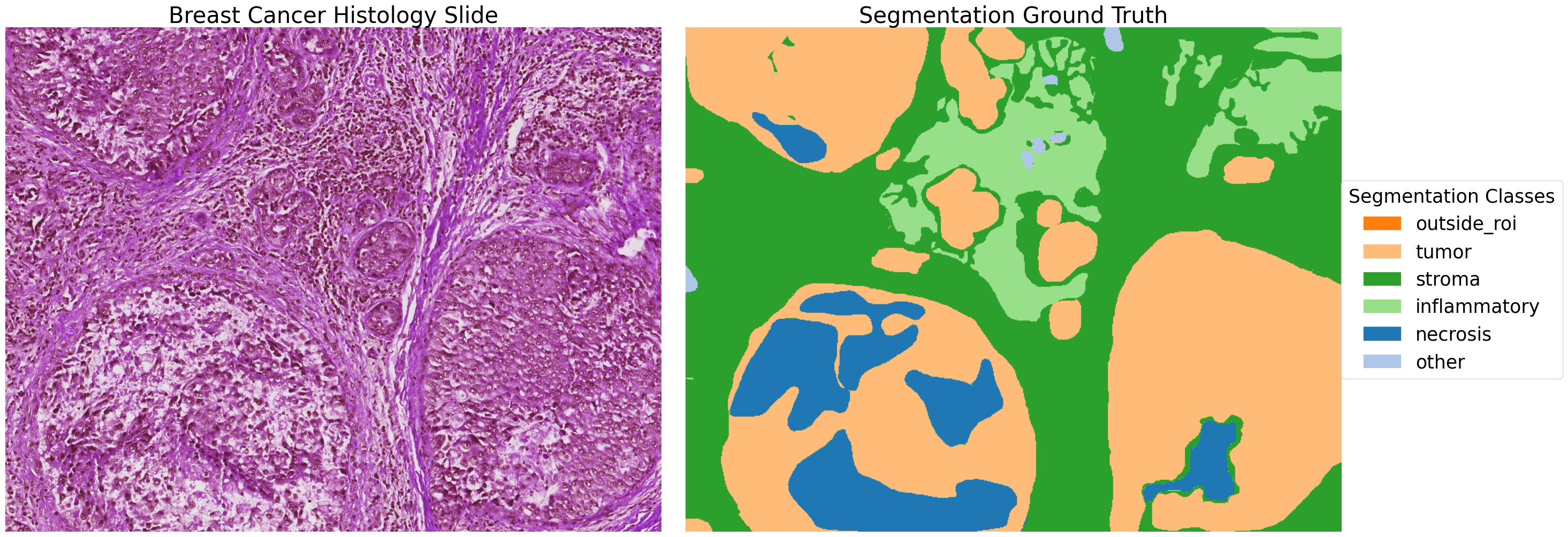}\hspace*{-1cm}
\vspace{-0.3cm}
\caption{Example patch from histology slide (left) and 6 class segmentation mask (right), with tumour cells shown in yellow}
\vspace{-1.0cm}
\end{figure}

In the original format, the image dataset \citep{BCSS_repo} contained 151 images of histology slides (each at resolution 0.25 microns/pixel) provided by a number of different laboratories, and 151 corresponding segmentation masks (both images and masks in .png image format). The dataset has a memory size of 4.59 GB, pixel widths ranging from 1222 to 9801 (mean of 4310) and pixel heights ranging from 957 to 9248 pixels (mean of 3629). Each image corresponds to a specific digitised histological slide \citep{amgad2019structured} (as shown in Figures \ref{figure:wsi_examples} and \ref{figure:seg_mask_examples}) and shows breast cancer cells and other biological tissues. Data is downloaded using code provided by the researchers \footnotemark{} \citep{BCSS_repo} and the data is prepared according to the methodology presented in the research paper that originally generated the dataset \citep{amgad2019structured}. 

To summarise the data preparation process, these steps involve collapsing the segmentation masks down from 22 classes to 6 classes, and then splitting each whole slide image (WSI) into overlapping patches of 800x800 pixels. The patch size of 800x800 was chosen to allow for direct comparisons of performance between models used by the researchers who initially generated the dataset and the results from models that we developed. 
\footnotetext{licensed under an MIT licence}

Each segmentation mask has the same height and width pixel dimensions as the image it corresponds to, however instead of the RGB pixel values present in the digitised slides, each segmentation mask contains an array of integer values where each integer signifies a certain class is present at that specific pixel location within the original image. Cancerous cells and healthy cells show certain visual differences and it is these differences that pathologists pay attention to when deciding if a ROI of the WSI shows cancerous cells. However, these visual differences can be subtle and a pathologist may need to spend a long period of time analysing a specific image or region of interest to correctly determine if cancerous cells are present. Following the methodology presented in the previous work \citep{amgad2019structured}, the initial provided ground truth masks of 22 classes were combined into 6 general classes as shown in Table \ref{class-grouping-table}.
\begin{table}[h]
\vspace*{-0.5cm}
  \caption{Segmentation mask class grouping}
  \label{class-grouping-table}
  \centering
  \begin{tabular*}{\linewidth}{c p{1.85cm} p{4.0cm}}
    \toprule
    \textbf{Class} & \textbf{Designation} & \textbf{Items included in class} \\
    \midrule
    0 & Out of scope & Outside region of interest \\
    \hline
    1 & Tumourous & Tumour, Angioinvasion, \newline Ductal carcinoma in situ \\
    \hline
    2 & Stroma & Stroma \\
    \hline
    3 & Inflammatory 
    infiltration & Inflammatory 
    infiltration,\newline
    Lymphocytes,\newline
    Plasma cells, \newline
    Other immune infiltrates \\
    \hline
    4 & Necrosis & Necrosis \\
    \hline
    5 & Other tissues & Other biological
    tissue from all other classes \\
    \bottomrule
  \end{tabular*}
  \vspace*{-1.0mm}
\end{table}
The large number of pixels present in WSIs makes it necessary to divide each image and its corresponding segmentation mask into overlapping 800x800 pixel patches, to facilitate easier loading into the machine learning model. This process yields a total of 4812 Region of Interest (ROI) patches, each patch comprising one RGB image and a corresponding mask. These ROIs are then distributed into training, validation, and test subsets as per standard machine learning methodologies. The allocation to subsets is based on the lab where the slide was scanned, following the methodology used when the dataset was first evaluated for segmentation tasks \citep{amgad2019structured,segment_baseline}. Images from labs coded as OL, LL, E2, EW, GM, and S3 were used in the original paper for model performance evaluation. It is unclear if the authors used a separate 'hold-out' testing set, so we chose 'E2' lab images as an independent test set. This allows us to better evaluate the model's performance on new unseen images after the completion of the research project and to simulate real-world clinical applications. The allocation of image/mask pairs to each subset is outlined in Table \ref{table:dataset_split} and the pixel level class distributions per subset are shown in Table \ref{table:pixel-level-distributions}. During a later stage of model comparisons, a number of different image augmentations were included in the model training process with the aim of making the machine learning model more generalisable and robust to the specific distribution of images within the training data. Finally k-fold cross validation is used to get an accurate measure of the models performance and to make sure the model is generalising as expected.

\begin{table}[h]
\vspace*{-0.5cm}
\caption{Training/validation/test image split}
\label{table:dataset_split}
\centering
    \begin{tabular*}{\linewidth}{lcc}
    \toprule
    \textbf{Data split} & \textbf{Patch count} & \textbf{\% of total} \\
    \midrule
    Training\footnotemark{} & 3318 & 0.691 \\
    Validation\footnotemark{} & 974 & 0.202 \\
    Testing\footnotemark{} & 520 & 0.108 \\
    \bottomrule
    \end{tabular*}
    \vspace*{-0.5cm}
\end{table}

\footnotetext{Training patch counts by lab -  E9:66, GI:15, HN:9, D8:621, BH:603, C8:261, A7:273, A8:88, AC:132, AN:150, AO:274, AQ:28, AR:204, A1:55, A2:539}
\footnotetext{Validation patch counts by lab - OL:188, LL:62, EW:441, GM:249, S3:34}
\footnotetext{Test patch counts by lab - E2:520}

\begin{table}[h]
\centering
\caption{Distribution of pixels within each data subset belonging to each specific class}
\vspace*{-0.2cm}
\label{table:pixel-level-distributions}
\begin{tabular*}{\linewidth}{cp{1.8cm}p{1.8cm}p{1.8cm}}
\toprule
\textbf{Class} & \textbf{Pixel dist. train (\%)} & 
\textbf{Pixel dist. val. (\%)} & 
\textbf{Pixel dist. test (\%)} \\
\midrule
0 & 15.32 & 33.74 & 28.55 \\
1 & 34.72 & 22.35 & 30.13 \\ 
2 & 29.22 & 27.34 & 26.54 \\ 
3 & 09.94 & 09.75 & 08.49 \\ 
4 & 06.27 & 02.60 & 03.78 \\ 
5 & 04.53 & 04.22 & 02.52 \\ 
\bottomrule
\end{tabular*}
\end{table}
\vspace{-0.3cm}
\subsection{General model training}
For each research question, a series of experiments were undertaken, assessing how changing the variable(s) of interest (e.g. model type, type, number and magnitude of image augmentations used, loss function etc.) would affect key metrics relating to the image segmentation task. The metrics chosen for the evaluation of each experiment are given as follows: Dice Score, IoU score (Jaccard Index), and Accuracy. 
\mycomment{
The formal definitions of these metrics are given below in Table \ref{table:metric_definitions}.
\begin{table}[h]
\caption{Metric definitions}
\label{table:metric_definitions}
\centering
\begin{tabular*}{\linewidth}{p{0.1\linewidth} l l l}
\toprule
\textbf{Metric} & 
\textbf{Set Notation} & \textbf{True Positive Notation} \\
\midrule
Dice & 
$\dfrac{{2 \cdot |A \cap B|}}{{|A| + |B|}}$ & $\dfrac{{2 \cdot TP}}{{2 \cdot TP + FP + FN}}$\\
IoU & 
$\dfrac{|A \cap B|}{|A| +  |B| - |{A} \cap {B}|}$ & $\dfrac{TP}{(TP + FP + FN)}$ \\
Accuracy & 
$\dfrac{|A \cap B| + |\bar{A} \cap \bar{B}|}{|A \cup B| + |\bar{A} \cup \bar{B}|}$ & $\dfrac{{TP + TN}}{{TP + TN + FP + FN}}$ \\
\bottomrule
\end{tabular*}
$A$ - predicted segmentation mask, $B$ - ground truth mask, $\overline{A}$ - predicted non-segmentation area,\\ $\overline{B}$ - ground truth non-segmentation area, TP - true positives, FN - false negatives,\\ FP - false positives, TN - true negatives
\end{table}
}
During model training, the performance of tumour class segmentation was assessed via Dice Score (for comparison against previous research \citep{amgad2019structured}) and mean IoU (mIoU) score (for comparison against another previous research \citep{lu2022automatic}). The accuracy metric was also included to highlight potential training issues, for example, high accuracy paired with a low Dice score for the tumour class, may suggest the model is over-fitting on the other non-tumour classes during the training process. Training and validation data splits, as defined in Table \ref{table:dataset_split}, were consistently maintained during model selection and augmentation comparisons in order to control for potential variation in metrics if training and validation sets were randomly generated before each run. It was only during the final evaluation phase that these data splits were modified and we combine the previous training and validation data and change to the k-fold cross validation methodology. 

We employed the mmsegmentation package\footnotemark{} \citep{mmseg_cite,mmseg_repo} for model training which is a PyTorch-based framework with a large selection of segmentation architectures and pretrained models. Model architectures, parameters, and hyperparameters are defined in a configuration file and are loaded into a 'runner' object which co-ordinates data loading, model training and model evaluation. Each model was trained in a virtual environment using a single Tesla G4 or V100 GPU (depending on GPU availability) and 15GB of GPU RAM. \footnotetext{licensed under the Apache 2.0 licence} In addition, the Albumentations package\footnotemark{} \citep{albu_paper,albu_repo} was also used as it contains a large number of image augmentation techniques (such as spatial transformations and colour manipulation). \footnotetext{licensed under the Apache 2.0 licence} The Optuna package\footnotemark{} \citep{optuna_paper,optuna_website} was used to search through the image augmentation search space and optimise image augmentation parameters such as number and magnitude of image augmentations.
\footnotetext{licensed under MIT license}

\subsection{Model architecture selection}
Informed by existing image segmentation literature, we chose to explore transformer-based models which have shown high performance on reference segmentation tasks like the ADE20K and Cityscapes datasets. The inherent attention mechanisms allow transformer-based segmentation models show improved performance over traditional CNN models by capturing whole image information and context \citep{li2023transforming}, which is crucial for accurate segmentation. Minimal image transformations/augmentations were used during model comparison experiments to control for the possible variation in model performance from random/probabilistic augmentations. Image resizing and adjustment of mean and standard deviation of image pixels on a per channel basis, facilitated effective transfer learning \citep{zhuang2021_trans_survey} from the pretrained models. We maintained a consistent training and validation split of images and masks during initial model comparison, where each model was trained for 1000 iterations. The one-cycle learning rate scheduler \citep{onecycle} was used with a maximum learning rate of 0.001, to prompt super-convergence whilst model training, providing a performance approximation for a larger number of training iterations. After 1000 iterations, we compared a model's segmentation predictions against the validation set, yielding per-class performance metrics for Dice score, IoU score, and accuracy.

The model architectures which were tested during the initial evaluation are given below:
\begin{center}
    SegFormer\footnotemark{} \citep{segformer}\\
    \footnotetext{Licensed under MIT license}
    Mask2Former\footnotemark{} \citep{mask2former}\\
    \footnotetext{Licensed under MIT license}
    PoolFormer\footnotemark{} \citep{poolformer}\\
    \footnotetext{Licensed under Apache 2.0 license}
    ConvNext\footnotemark{} \citep{convnext}\\ 
    \footnotetext{Licensed under MIT license}
    FCN\footnotemark{}
    \citep{FCN}\\
    \footnotetext{Unable to determine license conditions for this model}
\end{center}

As part of the secondary evaluation process, a number of models that showed promise during the initial stage were selected for further evaluation. During this stage, each model was trained for 2000 iterations, using cross-entropy loss and using the one-cycle learning rate scheduler, again set to a maximum learning rate of 0.001. We took into consideration model performance from the initial stage as well as performance against common segmentation reference datasets. 

\subsection{Image augmentation optimisation}

 We made use of the RandAugment methodology  for selecting image augmentations as this has been shown to give similar performance to more comprehensive augmentation search strategies but considerably narrows the complexity of the search space for possible image augmentation parameters \citep{randaugment}. A visualisation of some of the different image augmentations that were explored can be seen in Appendix  \ref{apd:augmentation_vis}. To search through the RandAugment search space of augmentations, we employed the Optuna package \citep{optuna_paper,optuna_website} and used a Tree-structured Parzen Estimator (TPE) algorithm \citep{bergstra2011algorithms} to find values of 'M' (relative strength of all image augmentations to be applied) and 'N' (number of augmentations to be applied) which would lead to a maximal Dice Score for segmenting the tumour class. Upper and lower limits for augmentation parameters are selected for each augmentation, and then between these lower and upper values, 30 values are selected which each correspond to a specific M value. During each optuna trial, values for M and N are generated, N augments are randomly selected (with equal probability) from K total image augmentations. Each trial, a model is trained using the randomly selected N augmentations applied to the input training data with an augmentation strength of M. For the next trial, M and N are adjusted via the Tree-based parzen estimator search optimiser based on the previous results in order to attempt to maximise the Dice Score for the tumour class. A baseline Dice Score is obtained by training a model ten times without augmentations and taking the mean of the Dice Scores for predictions against the validation set. The performance of each individual augmentation is evaluated by using the final value of 'M' found in the earlier augmentation optimisation stage, training a model five times, and taking the mean Dice Score. To streamline the augmentation search process, each model was trained for 500 epochs with a constant learning rate of 0.0001 and a resized patch size of 512x512. We selected the SegFormer\_B2 model for the optimisation process as it offers a good balance between training and inference speed, ensuring the optimisation process did not experience a bottleneck when the model makes predictions against the validation dataset. During the comparison of image augmentation techniques, we forced the model to use Automated Mixed Precision (AMP) during training, to speed up the computation of loss gradients and training times \citep{nvidia-mixed-precision}. Once values for M and N have been found that give the highest Dice Score, we then perform an ablation study to identify the impact of individual image augmentations. Ten trials without image augmentation are undertaken to obtain a mean Dice Score baseline measure and then for each augmentation in turn, only this single augmentation is applied, then a further ten trials are undertaken to obtain a mean Dice Score when this augmentation is applied. 

\subsection{Loss function comparison}
Loss functions play a crucial role in training machine learning models \citep{wang2022survey_loss}, particularly in image segmentation tasks \citep{jadon_loss_semantic}. Different loss functions can lead to variations in model behaviour during training. In general, loss metrics tailored to image segmentation, like Dice Score or Jaccard Index, tend to outperform more generic metrics like Cross-Entropy loss \citep{eelbode2020dice_jaccard}. To compare the different loss functions, the SegFormer\_B5 model was used. We trained the model for 1000 iterations, using a constant learning rate 0f 0.001, predicting against the validation set at 500 and 1000 iterations and taking the mean Dice Score to account for natural variations in Dice Score whilst training. 

\subsection{Combining best models and augmentations}

Once we have found the optimal values for M and N from the augmentation search process, we then examine the individual impact of the augmentations when each is applied with magnitude M. The N best-performing individual augmentations are selected. Next, we perform k-folds cross-validation with 10 folds. The training and validation images are combined into a single dataset, the image order is shuffled and then the images are randomly split into 10 folds. For each of the best-performing models, we perform two sets of k-fold cross-validation experiments. The first is to obtain a baseline for each model without augmentations. We train a particular model 10 different times (one for each fold) for 2000 epochs, and each time the model is trained, 9 of the folds are used for training and the 10th fold is used for validation. The Dice Score against the validation set is saved for each fold. Now we repeat the k-fold cross-validation process and this time, each time a model is trained, the best N augmentations with augmentation parameters of magnitude M are applied to the training data. The validation data during each fold is left in its original format without augmentations applied. 

\section{Discussion of results}

\subsection{Initial model evaluation}
In the primary evaluation phase, a number of different model architectures and subtypes are compared after being trained for 1000 iterations with the same split of training and validation images. A summary of the initial model comparison results is given in Table \ref{table:summary_init_model_performance} with models grouped by model type (detailed results can be found in the supplementary material).
\begin{table}[h]
\vspace*{-0.5cm}
\caption{Initial model comparison results, grouped by model type}
\label{table:summary_init_model_performance}
\begin{tabular*}{\linewidth}{l p{0.8cm} p{0.8cm} p{0.8cm} p{0.8cm}}
    \toprule
    \textbf{Model group} & \textbf{Mean Dice Score} & \textbf{Std Dice Score} & \textbf{Mean IoU} & \textbf{Std IoU} \\
    \midrule
   \textbf{Segformer} & 67.50 & 0.71 & 50.94 & 0.81\\
   Poolformer & 66.20 & 1.50 & 49.49 & 1.50 \\
   Mask2Former & 64.96	& 6.61  & 48.36 & 6.93 \\	
   ConvNext & 62.00 &	2.27 &	44.95 & 2.39 \\
    FCN	& 58.75 &	1.30 &	41.59 &	1.30 \\
    \bottomrule
  \end{tabular*}
\end{table}
Table \ref{table:summary_init_model_performance} shows summary statistics for the different model types during the initial model evaluation experiments, including Dice and IoU scores for predictions of tumour class segmentation compared against the validation dataset. As we see from the table, the transformer-based models show increased segmentation performance compared to the convolutional-based models. This is because transformer-based models can capture global context for a patch via the attention mechanism, allowing for dynamic focus on relevant image regions. Transformers can learn hierarchical representations which allows features at multiple scales to be captured, which is crucial for medical images where cellular structures can vary in size and shape. This makes transformers adept at handling the complexity and variability often seen in histology slides, providing more robust and accurate segmentation compared to traditional convolutional-based segmentation models.
\begin{table}[h]
\vspace*{-0.7cm}
\centering
\caption{Best performing models - initial evaluation}
\label{table:init_model_performance}
\begin{tabular*}{\linewidth}{lll}
\toprule
\textbf{Model name} &
\textbf{Val.Dice} & \textbf{Val.IoU} \\
\midrule
\textbf{Mask2former\_R51} &
69.03 & 52.76 \\
Poolformer\_FPN\_S24 &
68.31 & 51.87 \\
Mask2former\_Swin-T &
68.02 & 51.52 \\
Segformer\_B3 &
67.87 & 51.37 \\
Segformer\_B5 &
67.82 & 51.33 \\
Mask2former\_Swin-S &
67.73 & 51.21 \\
Poolformer\_FPN\_M36 &
66.41 & 49.71 \\
\bottomrule
\end{tabular*}
\end{table}

\subsection{Further model evaluation}
After considering the results from the initial evaluation, we decided not to continue experimenting with the ConvNext and FCN models due to their poor Dice Scores compared to the other types of models. We selected the following models for further evaluation: SegFormer\_B5, PoolFormer\_FPN\_S36, and PoolFormer\_FPN\_M36\footnotemark{} . and the results for these models are shown in Table \ref{table:model_further_evaluations}. \footnotetext{As part of the secondary evaluation process, it was planned to experiment further with the Mask2Former models such as M48 but we faced technical difficulties involving gradient instability and model divergence with the training process for this model type}
\begin{table}[h]
\vspace*{-0.5cm}
\caption{Results of the secondary evaluation experiments, showing Dice Score, IoU and accuracy after 2000 iterations}
\label{table:model_further_evaluations}
\centering
\begin{tabular*}{\linewidth}{p{4.0cm}lll}
\toprule
\textbf{Model Backbone} & 
\textbf{DS} & 
\textbf{IoU} & \textbf{Acc} \\
\midrule
\textbf{PoolFormer\_M36\_FPN} & 
70.27 & 
54.17 & 87.68 \\
PoolFormer\_S36\_FPN &
69.92 & 
53.75 & 89.99 \\
SegFormer\_B5 & 
68.84 & 
52.49 & 90.69 \\
\bottomrule
\end{tabular*}
\vspace*{-0.1cm}
\end{table}
We have shown that the PoolFormer model using an M48 backbone and an FPN decode head is able to achieve the best performance when segmenting the tumour class from histology slides.
\vspace*{-0.2cm}
\subsection{Comparison of loss types}
\begin{table}[h]
\vspace*{-0.6cm}
  \caption{Comparison of different loss types, showing Dice Scores at 500 and 1000 iterations and the mean Dice Score}
  \label{table:loss-comparison}
  \centering
  \begin{tabular*}{\linewidth}{llll}
    \toprule
    \textbf{Loss type} & \textbf{DS500} & \textbf{DS1000} & \textbf{Mean DS} \\
    \midrule
    Focal & 61.12 & \textbf{68.40} & \textbf{64.76} \\
    CE & 54.42 & 68.21 & 61.32 \\
    Lovasz & 47.99 & 63.49 & 55.74 \\
    Dice & \textbf{68.39} & 58.80 & 63.60 \\
    Tvarsky & 67.44 & 56.29 & 61.87 \\
    \bottomrule
  \end{tabular*}
  \vspace{-0.12cm}
\end{table}

It is shown in Table \ref{table:loss-comparison} that when trained for 1000 iterations with no image augmentations, the Segformer\_B5 model shows the best performance when using a Focal loss function, followed by Cross-Entropy loss function. As the mmsegmentation package allows multiple losses to be used during back-propagation, during the final evaluation of the models we attempted to create a custom loss combining Focal and CE losses. We reasoned that two loss functions working together would allow the model to generalise better as each loss function would penalise different sources of error. However, when attempting to train models using this custom loss function, models began to exhibit divergent behaviour and so we could not investigate fully the effect of combining these different loss types.

\subsection{Individual augmentation comparison}

After the image augmentation search, optimal RandAugment values were found to be N=10 and M=27 (see Appendix \ref{apd:RandAugment Search results} for a detailed table showing the progression of the N and M values during the optimisation process). Individual augmentations were then tested using this value of 'M' to set each set of augmentation parameters. Table \ref{table: Individual augmentations comparison} identifies the top five augmentations as Sharpen, GridDropout, Emboss, ColorJitter, and HueSaturationValue. ColorJitter and HueSaturationValue enhance colour variation in the training data, simulating staining differences between different labs when processing the histology slides. Sharpen and Emboss augmentations likely improve feature emphasis for cancerous cells by emphasising specific features of the cells and nucleus typically associated with cancerous cells, such as nucleus shape. GridDropout forces the model to consider adjacent image regions, enhancing visual context. GridDistortion and ElasticTransform augmentations are shown in previous research to give increased performance, yet here surprisingly they show decreased segmentation performance. This could be due to the high augmentation magnitude (M), causing excessive image distortion which impedes the training process.

\begin{table}[h]
\vspace*{-0.5cm}
\caption{Comparison of specific image augmentations, showing Dice Score percentage difference vs the baseline as well as the p-value of the difference in means between the baseline and augmentation trials}
  \label{table: Individual augmentations comparison}
  \centering
  \begin{tabular*}{\linewidth}{p{2.7cm}p{0.8cm}p{0.8cm}p{0.8cm}p{0.8cm}}
    \toprule
           \textbf{Augmentation} & \textbf{Mean DS} & 
           \textbf{\% Diff. in DS} &
        \textbf{Std DS} & 
        \textbf{p-val}\\
     & & \\
    \midrule
               \textbf{Sharpen} &  67.78 &                      1.51 &         0.34 &    0.023 \\
           GridDropout &  67.63 &                      1.29 &         0.29 &    0.045 \\
                Emboss &  67.52 &                      1.12 &         1.47 &    0.396 \\
           ColorJitter &  67.50 &                      1.09 &         0.91 &    0.240 \\
    HueSaturationValue &  67.37 &                      0.90 &         0.84 &    0.302 \\
     OpticalDistortion &  67.29 &                      0.78 &         1.07 &    0.440 \\
          GaussianBlur &  67.23 &                      0.69 &         1.47 &    0.591 \\
                CutOut &  66.84 &                      0.11 &         0.80 &    0.896 \\
                 CLAHE &  66.71 &                     -0.09 &         1.09 &    0.927 \\
        GridDistortion &  66.71 &                     -0.10 &         1.02 &    0.921 \\
        RandomContrast &  66.62 &                     -0.23 &         2.06 &    0.893 \\
      ElasticTransform &  65.88 &                     -1.34 &         1.14 &    0.223 \\
           RandomGamma &  60.28 &                     -9.72 &        11.33 &    0.316 \\
    \bottomrule
  \end{tabular*}
 \vspace*{-0.6cm}
\end{table}

\subsection{Combined best models and augmentation results}
 
\begin{table}[h]
\caption{Dice Score mean and standard deviations from k-fold cross-validation comparing baseline vs augmentations \protect\footnotemark and p-value for difference in means}
\label{table:k-folds-results}
\centering
\begin{tabular*}{\linewidth}{p{2.6cm}p{0.7cm}p{0.5cm}p{0.6cm}p{0.5cm}p{0.4cm}}
\toprule
\textbf{Model} & 
\textbf{BL Mean} & 
\textbf{BL Std} & 
\textbf{WA Mean} & \textbf{WA Std} & 
\textbf{p-val} \\
\midrule
\textbf{PoolFormer\_S36} & 81.07 & 1.37 & 81.05 & 1.56 & 0.98 \\
PoolFormer\_M36 & 80.84 & 1.35 & 81.10 & 1.14 & 0.36 \\
SegFormer\_B5 & 76.81 & 2.47 & 78.36 & 2.90 & 0.36 \\
\bottomrule
\end{tabular*}
\vspace*{-0.5cm}
\end{table}
The best metrics achieved during the k-fold cross-validation with augmentations were a Dice Score of 84.08 and an IoU score of 72.54 (achieved by the PoolFormer\_S36 model). A similar Dice Score is obtained compared to the closest work \citep{amgad2019structured} which achieved a Dice Score of 85.1 As indicated in Table \ref{table:k-folds-results}, when comparing Dice Scores, the image augmentation methodology did not yield any statistically significant improvements over the baseline models . Although initially, this appears to be a negative result, it offers a valuable insight: the Dice Scores remained stable even after the introduction of ten different image augmentations, applied at near maximum magnitude. Even though previous research \citep{lu2022automatic} uses a more technically advanced model architecture, we demonstrated a higher IoU than the value of 68.12 that the previous researchers achieved. This fact that the segmentation metrics are similar or better than existing research suggests that our methodology has resulted in models which have gained an enhanced ability to generalize across diverse input images and have become more robust to variations in input data, making it a versatile tool for segmenting tumorous cells in histology images from multiple labs without compromising segmentation performance. This resilience to data variability is a valuable attribute, especially in medical applications involving histology slides where data heterogeneity is often a challenge. Due to concerns regarding potential over-fitting in our model, due to the repeated number of model trainings that were undertaken during the augmentation optimisation process, we implemented several strategies to mitigate this risk. Notably, we incorporated weight decay, a form of L2 regularization, into our training process. This technique penalizes larger weights in the model, encouraging simpler models that are less likely to overfit. Furthermore, by employing the k-fold cross-validation methodology, this allows us to evaluate the model's performance across different subsets of the data. Throughout this process, we closely monitored the loss metrics for both training and validation sets. The absence of a significant increase in validation loss compared to training loss during any specific model training using a certain k-fold is a strong indicator of the model's generalization capabilities. 

\footnotetext{(BL: baseline without augmentations, WA: with augmentations, p-value calculated for difference in means with and without augmentations)}

\section{Limitations}
This study relies on the assumption of accurate segmentation masks in the dataset from previous research  \citep{amgad2019structured}. If these masks are inaccurate, the trained model's performance may be compromised. However, the segmentation masks in the dataset were validated by multiple pathology professionals which should ensure the annotations are accurate. Also, the images within the dataset are assumed to be representative of histology slides found in clinical settings. If the dataset is atypical in some manner, the model may fail to make accurate predictions on clinical images. The image augmentation optimisation process may cause overfitting on the validation set due to repeated predictions against the validation set. However, the increased variation of images within the training set after applying the augmentations should counterbalance this potential overfitting as well as the use of k-folds cross-validation during the final augmentation comparison process. Due to hardware and time constraints, we were not able to test our augmentation methodology in combination with larger models such as Beit \citep{bao2022beit} and due to technical issues during model training, we were not able to fully explore the performance of some Mask2Former models. It is possible that employing an auxiliary segmentation head, in combination with the best-performing models and augmentations may further increase segmentation performance. Training a model on Whole Slide Images (WSIs) rather than patched Regions of Interest (ROIs) could potentially improve segmentation performance, but it introduces additional challenges due to the large image sizes. Furthermore, testing the augmentation framework on models trained on a single dataset limits the generalisability of our findings. However, there are an extremely small number of annotated segmentation datasets relating to breast cancer histology slides so at this time it is not a trivial matter to expand our methodology to explore images from different datasets. Finally, employing model pretraining with histology-specific datasets \citep{kawai2023largescale} could also enhance baseline performance and the impact of the augmentation framework.
\vspace{-0.4cm}
\section{Conclusions}

In conclusion, this research set out to develop an innovative image augmentation methodology aimed at improving the performance of segmentation models for identifying tumorous cells in histology slides. The methodology yielded similar or better segmentation metrics than existing research and the methodology resulted in more robust models without suffering a loss of segmentation performance. This robustness to data variability is particularly crucial in the medical imaging domain, where data can differ substantially due to various factors such as equipment, preparation techniques, and human error. Our methodology, therefore, contributes to the broader goal of creating more reliable and versatile machine-learning models for medical applications, specifically for segmenting histology slides from different sources. Overall, this research offers valuable insights for the ongoing advancement of medical image segmentation technologies and the developed methodology and provided codebase will assist with the development of breast cancer cell segmentation from histology slides.

\section{Acknowledgements}
We would like to thank our colleagues at Cancer Research with AI (CARESAI) group for their assistance, support, and motivation in completing this research project. Additionally, we would like to give thanks to Lucas Puvis de Chavannes, Amina Chouigui, Abner Hernandez, Magnus Kallinger, and Andrii Krutsylo for their time spent reviewing the paper prior to submission and providing suggestions for improvements. 

\medskip
\small
\bibliography{breastcanc.bib}

\begin{thebibliography}{55}
\providecommand{\natexlab}[1]{#1}
\providecommand{\url}[1]{\texttt{#1}}
\expandafter\ifx\csname urlstyle\endcsname\relax
  \providecommand{\doi}[1]{doi: #1}\else
  \providecommand{\doi}{doi: \begingroup \urlstyle{rm}\Url}\fi

\bibitem[BCS()]{BCSS_repo}
Bcss github repository.
\newblock \url{https://github.com/PathologyDataScience/BCSS}.
\newblock Accessed: 2023-05-14.

\bibitem[nvi()]{nvidia-mixed-precision}
{NVIDIA Deep Learning Documentation - Mixed Precision Training}.
\newblock {NVIDIA Deep Learning Documentation}.
\newblock URL
  \url{https://docs.nvidia.com/deeplearning/performance/mixed-precision-training/index.html}.
\newblock Accessed: [13/05/2023].

\bibitem[pap()]{paperswithcode_vit}
Papers with code, vision transformers.
\newblock \url{'https://paperswithcode.com/method/vision-transformer}.
\newblock Accessed: [12/04/23].

\bibitem[seg()]{segment_baseline}
Bcseg2015 grand challenge - baseline.
\newblock \url{https://bcsegmentation.grand-challenge.org/Baseline/}.
\newblock Accessed on May 13, 2023.

\bibitem[Akiba et~al.(2019)Akiba, Sano, Yanase, Ohta, and Koyama]{optuna_paper}
Takuya Akiba, Shotaro Sano, Toshihiko Yanase, Takeru Ohta, and Masanori Koyama.
\newblock Optuna: A next-generation hyperparameter optimization framework.
\newblock In \emph{Proceedings of the 25th {ACM} {SIGKDD} International
  Conference on Knowledge Discovery and Data Mining}, 2019.

\bibitem[Aksac et~al.(2020)Aksac, Ozyer, Demetrick, et~al.]{aksac2020cactus}
A.~Aksac, T.~Ozyer, D.~J. Demetrick, et~al.
\newblock Cactus: cancer image annotating, calibrating, testing, understanding
  and sharing in breast cancer histopathology.
\newblock \emph{BMC Res Notes}, 13\penalty0 (1):\penalty0 13, 2020.
\newblock \doi{10.1186/s13104-019-4866-z}.
\newblock URL \url{https://doi.org/10.1186/s13104-019-4866-z}.

\bibitem[Albayrak and Bilgin(2018)]{albayrak2018automatic}
A.~Albayrak and G.~Bilgin.
\newblock Automatic cell segmentation in histopathological images via
  two-staged superpixel-based algorithms.
\newblock \emph{Medical \& Biological Engineering \& Computing}, 57\penalty0
  (3):\penalty0 653--665, Oct. 2018.
\newblock \doi{10.1007/S11517-018-1906-0}.

\bibitem[{Albumentations Team}(Accessed 2023-05-12)]{albu_repo}
{Albumentations Team}.
\newblock Albumentations: Fast and flexible image augmentations.
\newblock \url{https://github.com/albumentations-team/albumentations}, Accessed
  2023-05-12.

\bibitem[Alomar et~al.(2023)Alomar, Aysel, and Cai]{alomar2023data_aug}
Khaled Alomar, Halil~Ibrahim Aysel, and Xiaohao Cai.
\newblock Data augmentation in classification and segmentation: A survey and
  new strategies.
\newblock \emph{Journal of Imaging}, 9\penalty0 (2):\penalty0 46, 2023.
\newblock \doi{10.3390/jimaging9020046}.
\newblock URL \url{https://doi.org/10.3390/jimaging9020046}.

\bibitem[Amgad et~al.(2019)Amgad, Elfandy, Hussein, Atteya, Elsebaie, Elnasr,
  Sakr, Salem, Ismail, Saad, Ahmed, Elsebaie, Rahman, Ruhban, Elgazar, Alagha,
  Osman, Alhusseiny, Khalaf, Younes, Abdulkarim, Younes, Gadallah, Elkashash,
  Fala, Zaki, Beezley, Chittajallu, Manthey, Gutman, and
  Cooper]{amgad2019structured}
Mohamed Amgad, Habiba Elfandy, Hagar Hussein, Lamees~A Atteya, Mai A~T
  Elsebaie, Lamia S~Abo Elnasr, Rokia~A Sakr, Hazem S~E Salem, Ahmed~F Ismail,
  Anas~M Saad, Joumana Ahmed, Maha A~T Elsebaie, Mustafijur Rahman, Inas~A
  Ruhban, Nada~M Elgazar, Yahya Alagha, Mohamed~H Osman, Ahmed~M Alhusseiny,
  Mariam~M Khalaf, Abo-Alela~F Younes, Ali Abdulkarim, Duaa~M Younes, Ahmed~M
  Gadallah, Ahmad~M Elkashash, Salma~Y Fala, Basma~M Zaki, Jonathan Beezley,
  Deepak~R Chittajallu, David Manthey, David~A Gutman, and Lee A~D Cooper.
\newblock Structured crowdsourcing enables convolutional segmentation of
  histology images.
\newblock \emph{Bioinformatics}, 35\penalty0 (18):\penalty0 3461--3467, Sep
  2019.
\newblock \doi{10.1093/bioinformatics/btz083}.
\newblock URL \url{https://doi.org/10.1093/bioinformatics/btz083}.

\bibitem[Bancroft and Layton(2012)]{bancroft2012hematoxylins}
John~D Bancroft and Christopher Layton.
\newblock The hematoxylins and eosin.
\newblock \emph{Bancrofts theory and practice of histological techniques},
  7:\penalty0 173--186, 2012.

\bibitem[Bao et~al.(2022)Bao, Dong, Piao, and Wei]{bao2022beit}
Hangbo Bao, Li~Dong, Songhao Piao, and Furu Wei.
\newblock Beit: {BERT} pre-training of image transformers.
\newblock In \emph{International Conference on Learning Representations
  (ICLR)}, 2022.

\bibitem[Bergstra et~al.(2011)Bergstra, Bardenet, Bengio, and
  Kegl]{bergstra2011algorithms}
James Bergstra, Remi Bardenet, Yoshua Bengio, and Balazs Kegl.
\newblock Algorithms for hyper-parameter optimization.
\newblock In \emph{Conference on Neural Information Processing Systems
  (NeurIPS)}, 2011.

\bibitem[Buslaev et~al.(2020)Buslaev, Iglovikov, Khvedchenya, Parinov,
  Druzhinin, and Kalinin]{albu_paper}
Alexander Buslaev, Vladimir~I. Iglovikov, Eugene Khvedchenya, Alex Parinov,
  Mikhail Druzhinin, and Alexandr~A. Kalinin.
\newblock Albumentations: Fast and flexible image augmentations.
\newblock \emph{Information}, 11\penalty0 (2), 2020.
\newblock ISSN 2078-2489.
\newblock \doi{10.3390/info11020125}.
\newblock URL \url{https://www.mdpi.com/2078-2489/11/2/125}.

\bibitem[cancer.net()]{cancer_net}
cancer.net.
\newblock Breast cancer statistics.
\newblock \url{https://www.cancer.net/cancer-types/breast-cancer/statistics}.
\newblock Accessed: Date.

\bibitem[Cheng et~al.(2022)Cheng, Misra, Schwing, Kirillov, and
  Girdhar]{mask2former}
Bowen Cheng, Ishan Misra, Alexander~G. Schwing, Alexander Kirillov, and Rohit
  Girdhar.
\newblock Masked-attention mask transformer for universal image segmentation.
\newblock 2022.

\bibitem[Cohen(2020)]{cohen2020artificial}
S.~Cohen.
\newblock \emph{Artificial Intelligence and Deep Learning in Pathology}.
\newblock Elsevier Health Sciences, 2020.

\bibitem[Contributors(2020)]{mmseg_cite}
MMSegmentation Contributors.
\newblock Openmmlab semantic segmentation toolbox and benchmark.
\newblock \url{https://github.com/open-mmlab/mmsegmentation}, July 2020.
\newblock License: Apache-2.0.

\bibitem[Cubuk et~al.(2020)Cubuk, Zoph, Shlens, and Le]{randaugment}
Ekin~D. Cubuk, Barret Zoph, Jonathon Shlens, and Quoc~V. Le.
\newblock Randaugment: Practical automated data augmentation with a reduced
  search space.
\newblock \emph{Computer Vision and Pattern Recognition - cs.CV conference},
  2020.

\bibitem[Das et~al.(2020)Das, Nair, and Peter]{das2020computer}
Asha Das, Madhu~S Nair, and S~David Peter.
\newblock Computer-aided histopathological image analysis techniques for
  automated nuclear atypia scoring of breast cancer: a review.
\newblock \emph{Journal of digital imaging}, 33:\penalty0 1091--1121, 2020.

\bibitem[Demir and Yener(2005)]{demir2005automated}
C.~Demir and B.~Yener.
\newblock Automated cancer diagnosis based on histopathological images: a
  systematic survey.
\newblock Technical report, Rensselaer Polytechnic Institute, 2005.

\bibitem[Eelbode et~al.(2020)Eelbode, Bertels, Berman, Vandermeulen, Maes,
  Bisschops, and Blaschko]{eelbode2020dice_jaccard}
Tom Eelbode, Jeroen Bertels, Maxim Berman, Dirk Vandermeulen, Frederik Maes,
  Raf Bisschops, and Matthew~B. Blaschko.
\newblock Optimization for medical image segmentation: Theory and practice when
  evaluating with dice score or jaccard index.
\newblock \emph{IEEE Transactions on Medical Imaging}, 39\penalty0
  (11):\penalty0 3679--3690, 2020.
\newblock \doi{10.1109/TMI.2020.3002417}.

\bibitem[Han et~al.(2023)Han, Wang, Chen, Chen, Guo, Liu, Tang, Xiao, Xu, Xu,
  Yang, Zhang, and Tao]{han2023survey_vision}
Kai Han, Yunhe Wang, Hanting Chen, Xinghao Chen, Jianyuan Guo, Zhenhua Liu,
  Yehui Tang, An~Xiao, Chunjing Xu, Yixing Xu, Zhaohui Yang, Yiman Zhang, and
  Dacheng Tao.
\newblock A survey on vision transformer.
\newblock \emph{IEEE Transactions on Pattern Analysis and Machine
  Intelligence}, 45\penalty0 (1):\penalty0 87--110, 2023.
\newblock \doi{10.1109/TPAMI.2022.3152247}.

\bibitem[He et~al.(2012)He, Long, Antani, and Thoma]{he2012histology}
L.~He, L.R. Long, S.~Antani, and GR. Thoma.
\newblock Histology image analysis for carcinoma detection and grading.
\newblock \emph{Computer Methods and Programs in Biomedicine}, 107\penalty0
  (3):\penalty0 538--556, 2012.

\bibitem[Holzinger et~al.(2020)Holzinger, Goebel, Mengel, and
  Müller]{holzinger2020artificial}
A.~Holzinger, R.~Goebel, M.~Mengel, and H.~Müller.
\newblock \emph{Artificial Intelligence and Machine Learning for Digital
  Pathology: State-of-the-art and Future Challenges}, volume 12090.
\newblock Springer Nature, 2020.

\bibitem[Jadon(2020)]{jadon_loss_semantic}
Shruti Jadon.
\newblock A survey of loss functions for semantic segmentation.
\newblock In \emph{2020 IEEE Conference on Computational Intelligence in
  Bioinformatics and Computational Biology (CIBCB)}, pages 1--7, 2020.
\newblock \doi{10.1109/CIBCB48159.2020.9277638}.

\bibitem[Jia et~al.(2017)Jia, Huang, Chang, and Xu]{jia2017constrained}
Z.~Jia, X.~Huang, E.~I.-C. Chang, and Y.~Xu.
\newblock Constrained deep weak supervision for histopathology image
  segmentation.
\newblock \emph{IEEE Transactions on Medical Imaging}, 36\penalty0
  (11):\penalty0 2376--2388, Jul. 2017.
\newblock \doi{10.1109/TMI.2017.2724070}.

\bibitem[Kawai et~al.(2023)Kawai, Ota, and Yamaoka]{kawai2023largescale}
Masataka Kawai, Noriaki Ota, and Shinsuke Yamaoka.
\newblock Large-scale pretraining on pathological images for fine-tuning of
  small pathological benchmarks, 2023.

\bibitem[Krithiga and Geetha(2020)]{krithiga2020breast}
R.~Krithiga and P.~Geetha.
\newblock Breast cancer detection, segmentation and classification on
  histopathology images analysis: A systematic review.
\newblock \emph{Archives of Computational Methods in Engineering}, 28\penalty0
  (4):\penalty0 2607--2619, Aug. 2020.
\newblock \doi{10.1007/S11831-020-09470-W}.

\bibitem[Kumar et~al.(2017)Kumar, Abbas, and Aster]{kumar2017robbins}
V.~Kumar, A.K. Abbas, and J.C. Aster.
\newblock \emph{Robbins Basic Pathology E-book}.
\newblock Elsevier Health Sciences, 2017.

\bibitem[Lakshmanan and Saravanakumar(2018)]{lakshmanan2018nucleus}
B.~Lakshmanan and S.~Saravanakumar.
\newblock Nucleus segmentation in breast histopathology images.
\newblock In \emph{2018 International Conference on Communication and
  Computational Technologies (ICCTCT)}, Mar. 2018.
\newblock \doi{10.1109/ICCTCT.2018.8550929}.

\bibitem[Li et~al.(2023)Li, Chen, Tang, Wang, Landman, and
  Zhou]{li2023transforming}
Jun Li, Junyu Chen, Yucheng Tang, Ce~Wang, Bennett~A Landman, and S~Kevin Zhou.
\newblock Transforming medical imaging with transformers? a comparative review
  of key properties, current progresses, and future perspectives.
\newblock \emph{Medical image analysis}, page 102762, 2023.

\bibitem[Liu et~al.(2022)Liu, Mao, Wu, Feichtenhofer, Darrell, and
  Xie]{convnext}
Zhuang Liu, Hanzi Mao, Chao-Yuan Wu, Christoph Feichtenhofer, Trevor Darrell,
  and Saining Xie.
\newblock A convnet for the 2020s.
\newblock \emph{Proceedings of the IEEE/CVF Conference on Computer Vision and
  Pattern Recognition (CVPR)}, 2022.

\bibitem[Long et~al.(2015)Long, Shelhamer, and Darrell]{FCN}
Jonathan Long, Evan Shelhamer, and Trevor Darrell.
\newblock Fully convolutional networks for semantic segmentation.
\newblock In \emph{Conference on Computer Vision and Pattern Recognition
  (CVPR)}. IEEE, 2015.

\bibitem[Lu and Zhu(2022)]{lu2022automatic}
Xuan Lu and Xiaofeng Zhu.
\newblock Automatic segmentation of breast cancer histological images based on
  dual-path feature extraction network.
\newblock \emph{Mathematical Biosciences and Engineering}, 19\penalty0
  (11):\penalty0 11137--11153, Aug 2022.
\newblock \doi{10.3934/mbe.2022519}.

\bibitem[Madabhushi and Lee(2016)]{madabhushi2016image}
A.~Madabhushi and G.~Lee.
\newblock Image analysis and machine learning in digital pathology: Challenges
  and opportunities.
\newblock \emph{Medical Image Analysis}, 33:\penalty0 170--175, 2016.

\bibitem[Mills et~al.(2012)Mills, Carter, Greenson, Reuter, and
  Stoler]{mills2012sternberg}
S.E. Mills, D.~Carter, J.K. Greenson, V.E. Reuter, and M.H. Stoler.
\newblock \emph{Sternberg's Diagnostic Surgical Pathology}.
\newblock Lippincott Williams \& Wilkins, 2012.

\bibitem[Minaee et~al.(2022{\natexlab{a}})Minaee, Boykov, Porikli, Plaza,
  Kehtarnavaz, and Terzopoulos]{minaee2022image}
S.~Minaee, Y.~Boykov, F.~Porikli, A.~Plaza, N.~Kehtarnavaz, and D.~Terzopoulos.
\newblock Image segmentation using deep learning: A survey.
\newblock \emph{IEEE Transactions on Pattern Analysis and Machine
  Intelligence}, 44\penalty0 (7):\penalty0 3523--3542, Jun. 2022{\natexlab{a}}.
\newblock \doi{10.1109/TPAMI.2021.3051370}.

\bibitem[Minaee et~al.(2022{\natexlab{b}})Minaee, Boykov, Porikli, Plaza,
  Kehtarnavaz, and Terzopoulos]{minaee2022seg_review}
Shervin Minaee, Yuri Boykov, Fatih Porikli, Antonio Plaza, Nasser Kehtarnavaz,
  and Demetri Terzopoulos.
\newblock Image segmentation using deep learning: A survey.
\newblock \emph{IEEE Transactions on Pattern Analysis and Machine
  Intelligence}, 44\penalty0 (7):\penalty0 3523--3542, 2022{\natexlab{b}}.
\newblock \doi{10.1109/TPAMI.2021.3059968}.

\bibitem[Moorthy and Gandhi(2022)]{moorthy2022medsurvey}
Jayashree Moorthy and Usha~Devi Gandhi.
\newblock A survey on medical image segmentation based on deep learning
  techniques.
\newblock \emph{Big Data and Cognitive Computing}, 6\penalty0 (4), 2022.
\newblock ISSN 2504-2289.
\newblock URL \url{https://www.mdpi.com/2504-2289/6/4/117}.

\bibitem[{OpenMMLab}(Accessed 2023-05-12)]{mmseg_repo}
{OpenMMLab}.
\newblock Openmmlab semantic segmentation toolbox.
\newblock \url{https://github.com/open-mmlab/mmsegmentation}, Accessed
  2023-05-12.

\bibitem[{Optuna Contributors}(Accessed 2023-05-12)]{optuna_website}
{Optuna Contributors}.
\newblock Optuna: A hyperparameter optimization framework.
\newblock \url{https://optuna.org/}, Accessed 2023-05-12.

\bibitem[Ott et~al.(2009)Ott, Ullrich, and Miller]{ott2009importance}
J.~J. Ott, A.~Ullrich, and A.~B. Miller.
\newblock The importance of early symptom recognition in the context of early
  detection and cancer survival.
\newblock \emph{European Journal of Cancer}, 45\penalty0 (16):\penalty0
  2743--2748, Sep. 2009.
\newblock \doi{10.1016/j.ejca.2009.07.027}.

\bibitem[Rashmi et~al.(2021)Rashmi, Prasad, and Udupa]{rashmi2021breast}
R.~Rashmi, K.~Prasad, and CBK. Udupa.
\newblock Breast histopathological image analysis using image processing
  techniques for diagnostic purposes: A methodological review.
\newblock \emph{J Med Syst}, 46\penalty0 (1):\penalty0 7, Dec 2021.
\newblock \doi{10.1007/s10916-021-01786-9}.

\bibitem[Roy and Gupta(2020)]{roy2020macroscopic}
Bijoyeta Roy and Mousumi Gupta.
\newblock Macroscopic reconstruction for histopathology images: A survey.
\newblock In \emph{Computer Vision and Machine Intelligence in Medical Image
  Analysis: International Symposium, ISCMM 2019}, pages 101--112. Springer,
  2020.

\bibitem[Sayed et~al.(2015)Sayed, Lukande, and Fleming]{sayed2015providing}
S.~Sayed, R.~Lukande, and K.~A. Fleming.
\newblock Providing pathology support in low-income countries.
\newblock \emph{JCO Global Oncology}, 1\penalty0 (1):\penalty0 3--6, Sep. 2015.
\newblock \doi{10.1200/JGO.2015.000547}.

\bibitem[Smith and Topin(2018)]{onecycle}
Leslie~N. Smith and Nicholay Topin.
\newblock Super-convergence: Very fast training of neural networks using large
  learning rates, 2018.

\bibitem[Tellez et~al.(2019)Tellez, Litjens, Bándi, Bulten, Bokhorst, Ciompi,
  and {van der Laak}]{tellez2019_data_aug_hist}
David Tellez, Geert Litjens, Péter Bándi, Wouter Bulten, John-Melle Bokhorst,
  Francesco Ciompi, and Jeroen {van der Laak}.
\newblock Quantifying the effects of data augmentation and stain color
  normalization in convolutional neural networks for computational pathology.
\newblock \emph{Medical Image Analysis}, 58:\penalty0 101544, 2019.
\newblock ISSN 1361-8415.
\newblock \doi{https://doi.org/10.1016/j.media.2019.101544}.
\newblock URL
  \url{https://www.sciencedirect.com/science/article/pii/S1361841519300799}.

\bibitem[Tosta et~al.(2017)Tosta, Neves, and Nascimento]{tosta2017segmentation}
T.~A.~A. Tosta, L.~A. Neves, and M.~Z.~do Nascimento.
\newblock Segmentation methods of h\&e-stained histological images of lymphoma:
  A review.
\newblock \emph{Informatics in Medicine Unlocked}, 9:\penalty0 35--43, Jan.
  2017.
\newblock \doi{10.1016/J.IMU.2017.05.009}.

\bibitem[Wang et~al.(2022)Wang, Ma, Zhao, and Tian]{wang2022survey_loss}
Qi~Wang, Yue Ma, Kun Zhao, and Yingjie Tian.
\newblock A comprehensive survey of loss functions in machine learning.
\newblock \emph{Annals of Data Science}, 9\penalty0 (2):\penalty0 187--212,
  April 2022.
\newblock \doi{10.1007/s40745-020-00253-5}.

\bibitem[Xie et~al.(2021)Xie, Wang, Yu, Anandkumar, Alvarez, and
  Luo]{segformer}
Enze Xie, Wenhai Wang, Zhiding Yu, Anima Anandkumar, Jose~M. Alvarez, and Ping
  Luo.
\newblock Segformer: Simple and efficient design for semantic segmentation with
  transformers.
\newblock In \emph{Proceedings of the 35th Conference on Neural Information
  Processing Systems (NeurIPS)}, 2021.

\bibitem[Yan et~al.(2020)Yan, Ren, Wang, Wang, Zhang, Liu, Rao, Zheng, and
  Zhang]{yan2020breast}
Rui Yan, Fei Ren, Zihao Wang, Lihua Wang, Tong Zhang, Yudong Liu, Xiaosong Rao,
  Chunhou Zheng, and Fa~Zhang.
\newblock Breast cancer histopathological image classification using a hybrid
  deep neural network.
\newblock \emph{Methods}, 173:\penalty0 52--60, 2020.

\bibitem[Yu et~al.(2022)Yu, Luo, Zhou, Si, Zhou, Wang, Feng, and
  Yan]{poolformer}
Weihao Yu, Mi~Luo, Pan Zhou, Chenyang Si, Yichen Zhou, Xinchao Wang, Jiashi
  Feng, and Shuicheng Yan.
\newblock Metaformer is actually what you need for vision.
\newblock In \emph{Proceedings of the IEEE/CVF Conference on Computer Vision
  and Pattern Recognition}, pages 10819--10829, 2022.

\bibitem[Zhuang et~al.(2021)Zhuang, Qi, Duan, Xi, Zhu, Zhu, Xiong, and
  He]{zhuang2021_trans_survey}
Fuzhen Zhuang, Zhiyuan Qi, Keyu Duan, Dongbo Xi, Yongchun Zhu, Hengshu Zhu, Hui
  Xiong, and Qing He.
\newblock A comprehensive survey on transfer learning.
\newblock \emph{Proceedings of the IEEE}, 109\penalty0 (1):\penalty0 43--76,
  2021.
\newblock \doi{10.1109/JPROC.2020.3004555}.

\bibitem[Ziegenhorn and et~al.(2020)]{ziegenhorn2020breast}
H.-V. Ziegenhorn and et~al.
\newblock Breast cancer pathology services in sub-saharan africa: a survey
  within population-based cancer registries.
\newblock \emph{BMC Health Services Research}, 20\penalty0 (1):\penalty0 1--10,
  Oct. 2020.
\newblock \doi{10.1186/s12913-020-05726-0}.

\end{thebibliography}

\newpage
\section{Appendices}
\appendix

\section{Visualisation of augmentations}\label{apd:augmentation_vis}
\begin{figure}[h!]
\begin{center}
\setlength{\belowcaptionskip}{-2pt}
\label{figure:augmentation_vis}
\includegraphics[width=1.0\linewidth,height=18cm,keepaspectratio]{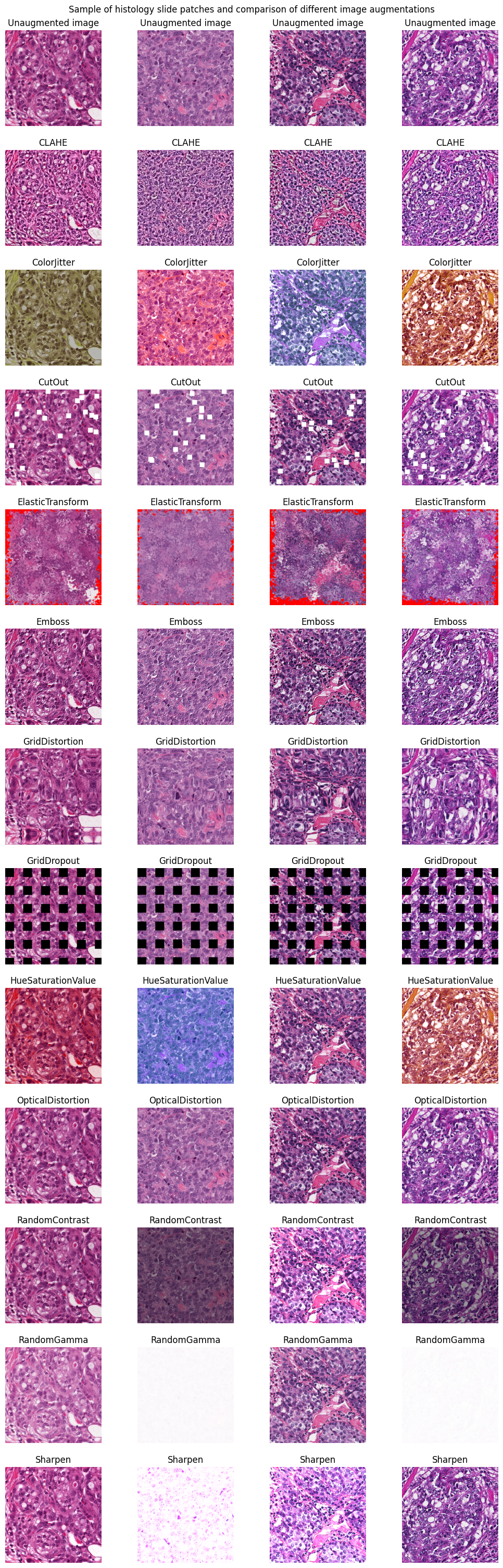}
\end{center}
\end{figure}

\section{RandAugment Search results}\label{apd:RandAugment Search results}
Dice Scores against the validation set, found during the search process for different RandAugment values of M and N

\begin{table}[h!]
\label{table:final_scores}
  \begin{tabular}{ccc}
    \toprule
    \textbf{M} & \textbf{N} & \textbf{Final Dice Score} \\
    \midrule
    27 & 9 & 69.30 \\
    21 & 5 & 69.29 \\
    21 & 3 & 68.90 \\
    9 & 2 & 68.38 \\
    21 & 7 & 68.03 \\
    17 & 6 & 68.00 \\
    15 & 4 & 67.93 \\
    25 & 6 & 67.61 \\
    25 & 7 & 67.53 \\
    13 & 3 & 67.32 \\
    21 & 10 & 67.04 \\
    23 & 5 & 66.83 \\
    13 & 1 & 66.37 \\
    29 & 8 & 66.32 \\
    29 & 8 & 66.15 \\
    27 & 9 & 66.11 \\
    19 & 5 & 65.71 \\
    7 & 10 & 63.74 \\
    25 & 4 & 63.46 \\
    \bottomrule
  \end{tabular}
\end{table}

\section{RandAugment augmentation parameter generation corresponding to various M values}\label{apd:first}

Lower and upper limits for the augmentation parameters were chosen from reviewing the existing literature and looking for commonly used values. For each parameter, 30 values were generated between the lower and upper limits (inclusive). These values were rounded to the nearest integer if the augmentation parameter is unable to accept floats as inputs. 

\begin{verbatim}
augmentation_dict = {
'RandomGamma': {
'gamma_limit_upper':
equally_spaced_nums(0, 120, 30,True)
},
'CLAHE': {
'clip_limit': 
equally_spaced_nums(0, 10, 30, True),
'tile_grid_size': 
equally_spaced_nums(8, 64, 30, True)
},
'CutOut': {
'num_holes':
equally_spaced_nums(0, 30, 30, True),
'max_h_size':
equally_spaced_nums(0, 20, 30, True),
'max_w_size': 
equally_spaced_nums(0, 20, 30, True)
},
'HueSaturationValue': {
'hue_shift_limit':
equally_spaced_nums(0, 50, 30, True),
'sat_shift_limit':
equally_spaced_nums(0, 50, 30, True),
'val_shift_limit':
equally_spaced_nums(0, 50, 30, True)
},
'ColorJitter': {
'brightness': equally_spaced_nums
(0, 1.0, 30, False),
'contrast': 
equally_spaced_nums (0, 1.0, 30, False),
'saturation':
equally_spaced_nums(0, 1.0, 30, False),
'hue': 
equally_spaced_nums(0, 1.0, 30, False)
},
'ElasticTransform': {
'alpha': 
equally_spaced_nums(0, 1000, 30, True),
'sigma': 
equally_spaced_nums(0, 50, 30, True),
'alpha_affine':
equally_spaced_nums(0, 60, 30, True)},
'OpticalDistortion': {
'distort_limit_high':
equally_spaced_nums(0, 0.5, 30, True),
'shift_limit_high':
equally_spaced_nums(0, 0.5, 30, True)
},
'GridDistortion': {
'num_steps':
equally_spaced_nums(4, 20, 30, True),
'distort_limit':
equally_spaced_nums(0, 1.0, 30, False)
},
'RandomContrast': {
'contrast_limit_upper':
equally_spaced_nums(0, 1.0, 30, False)
},
'GridDropout': {
'unit_size_min':
equally_spaced_nums(10, 210, 30, True),
'unit_size_max':
equally_spaced_nums(10, 210, 30, True),
'holes_num':
equally_spaced_nums(0, 100, 30, True),
},
'GaussianBlur': {
'blur_limit':
equally_spaced_nums(0, 10, 30, False),
'sigma_limit':
equally_spaced_nums(0.0, 5.0, 30, False)
},
'Sharpen': {
'alpha': equally_spaced_nums(0.0, 1.0, 30, False),
'lightness': equally_spaced_nums(0, 10, 30, False)
},
'Emboss': {
'alpha': equally_spaced_nums(0.0, 1.0, 30, False),
'strength': equally_spaced_nums(0.0, 2.0, 30, False)
}
}
\end{verbatim}

\section{Final augmentations and parameters obtained from augmentation search}

\begin{verbatim}
[{'type': 'CLAHE',
'clip_limit': 9,
'tile_grid_size': 60, 
'p': 1.0}, 
{'type': 'ColorJitter',
'brightness': 0.931,
'contrast': 0.931,
'saturation': 0.931, 
'hue': 0.931, 'p': 1.0}, 
{'type': 'CutOut',
'num_holes': 28,
'max_h_size': 19,
'max_w_size': 19,
'fill_value': 255, 
'p': 1.0}, 
{'type': 'Emboss', 
'alpha': 0.931, 
'strength': 1.8621, 
'p': 1.0}, 
{'type': 'GaussianBlur',
'blur_limit': 9.3103,
'sigma_limit': 4.6552, 
'p': 1.0}, 
{'type': 'GridDropout',
'ratio': 0.5,
'unit_size_min': 0,
'unit_size_max': 196, 
'holes_number_x': 93,
'holes_number_y': 93,
'shift_x': 0,
'shift_y': 0, 
'random_offset': False,
'fill_value': 0,
'mask_fill_value': None,
'always_apply': False, 'p': 1.0}, 
{'type': 'HueSaturationValue',
'hue_shift_limit': 47,
'sat_shift_limit': 47,
'val_shift_limit': 47,
'p': 1.0}, 
{'type': 'OpticalDistortion',
'distort_limit': 0.2,
'shift_limit': 0.2, 'p': 1.0}, 
{'type': 'Sharpen',
'alpha': 0.931,
'lightness': 9.3103,
'p': 1.0}]
\end{verbatim}


\end{document}